\documentclass[aps,prl,twocolumn,showpacs,superscriptaddress]{revtex4}

\usepackage{epsfig,amsmath,amsfonts,amssymb,psfrag}
\usepackage{color}

\begin{document}

\title{Mesoscopic Shelving Readout of Superconducting Qubits in Circuit QED}

\date{\today}

\author{B.G.U. Englert}
\thanks{Current address: Max-Planck-Institut f\"{u}r Quantenoptik, Hans-Kopfermann-Str. 1,
               D-85748 Garching, Germany.}
\affiliation{Walther-Mei{\ss}ner-Institut, Bayerische Akademie der Wissenschaften,
            Walther-Mei{\ss}ner-Str.~8, D-85748 Garching, Germany}
\affiliation{Physik Department, ASC and CeNS, Ludwig-Maximilians-Universit\"{a}t,
            Theresienstr.~37, D-80333 M\"{u}nchen, Germany}
\affiliation{Physik Department, Technische Universit\"{a}t
             M\"{u}nchen, D-85748 Garching, Germany}
\author{G. Mangano}
\affiliation{MATIS-INFM $\&$ Dipartimento di Metodologie Fisiche e Chimiche (DMFCI),
            viale A. Doria 6, 95125 Catania, Italy}
\affiliation{Institut f\"ur Theoretische Physik,
            Universit\"at Regensburg, D-93040 Regensburg, Germany}
\author{M. Mariantoni}
\affiliation{Walther-Mei{\ss}ner-Institut, Bayerische Akademie der Wissenschaften,
            Walther-Mei{\ss}ner-Str.~8, D-85748 Garching, Germany}
\affiliation{Physik Department, Technische Universit\"{a}t
             M\"{u}nchen, D-85748 Garching, Germany}
\author{R. Gross}
\affiliation{Walther-Mei{\ss}ner-Institut, Bayerische Akademie der Wissenschaften,
            Walther-Mei{\ss}ner-Str.~8, D-85748 Garching, Germany}
\affiliation{Physik Department, Technische Universit\"{a}t
             M\"{u}nchen, D-85748 Garching, Germany}
\author{J. Siewert}
\affiliation{Institut f\"ur Theoretische Physik,
            Universit\"at Regensburg, D-93040 Regensburg, Germany}
\author{E. Solano}
\affiliation{Departamento de Qu\'{\i}mica F\'{\i}sica, Universidad del Pa\'{\i}s Vasco -
            Euskal Herriko Unibertsitatea, Apdo. 644, 48080 Bilbao, Spain}

\begin{abstract}
We present a method for measuring the internal state of a
superconducting qubit inside an on-chip microwave resonator. We show
that one qubit state can be associated with the generation of an
increasingly large cavity coherent field, while the other remains
associated with the vacuum. By measuring the outgoing resonator field
with conventional devices, an efficient single-shot QND-like qubit readout can be achieved, enabling a high-fidelity
measurement in the spirit of the electron-shelving technique for
trapped ions. We expect that the proposed ideas can be adapted
to different superconducting qubit designs and contribute to
the further improvement of qubit readout fidelity.
\end{abstract}

\maketitle

Superconducting nanocircuits~\cite{makhlin01,qubitpapers} are
considered promising candidates for diverse implementations of
quantum information tasks~\cite{bouwmeester08}. In this context,
circuit quantum electrodynamics
(QED)~\cite{blais04,chiorescuwallraff}, which studies
superconducting qubits~\cite{makhlin01,nakamura99} coupled to
on-chip microwave resonators, occupies a central
role. To achieve the desired goals, it is important to implement
high-fidelity two-qubit gates~\cite{gatepapers} and efficient
schemes to read out the qubit state~\cite{readoutpapers}. In both
cases, trapped-ion systems represent the
state-of-the-art for qubit realizations~\cite{leibfried03}. In
particular, electron-shelving qubit readout has produced fidelity
benchmarks of approximately $99.99\%$~\cite{ionpapers}. These
astonishing achievements suggest the potential impact of
transferring key ideas from quantum optics to circuit QED.
Unfortunately, electron shelving relies strongly on the use of
single-photon detectors~\cite{leibfried03}, which are unavailable in
microwave technology in the range
$1-10\,\rm{GHz}$~\cite{romero08}. Nevertheless, in this manuscript
we show that a single-shot QND-like fast qubit readout can
be designed by exploiting the electron-shelving concept in circuit QED.

\begin{figure}[t]
\centering
\includegraphics[width=0.95\linewidth]{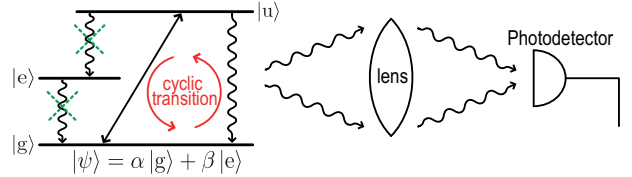}
\caption{(Color online) Sketch of electron shelving in trapped ions.
The $| {\rm g} \rangle \leftrightarrow | {\rm u} \rangle$ transition is driven with a laser beam, performing a cyclic transition and emitting many photons when $| {\rm g} \rangle$ is projected. No photons are detected when $| {\rm e} \rangle$ is measured.
Undesired transitions are inhibited via selection rules.}
\label{shelvingions}
\end{figure}

We first present the physics of electron shelving in
trapped ions. In Fig.~\ref{shelvingions}, we show a three-level atom where an
unknown qubit state $| \psi \rangle = \alpha | {\rm g} \rangle +
\beta | {\rm e} \rangle$ is encoded in states
$| {\rm g} \rangle$ and $| {\rm e} \rangle$. Via a laser beam, the
ground state $| {\rm g} \rangle$ is coupled to a third level $|
{\rm u} \rangle$, which can decay producing a continuous
cyclic transition. In this case, the qubit is projected onto state
$| {\rm g} \rangle$ and many photons are emitted in free space,
one at each cycle. In contrast, when the qubit is projected onto
state $| {\rm e} \rangle$, no photons are emitted. A lens is used
to collect the photons more efficiently by improving the solid
angle. Although the photodetector has a low efficiency $\eta_{\rm d}$, the qubit readout
fidelity can be very high. Typically, it is estimated through $F =
1 - e^{- \eta_{\rm d} N}$, which rapidly approaches unity for
$\eta_{\rm d} N \gg 1$, $N$ being the number of detected photons. Key elements
for electron shelving are the use of
three-level qubits, cyclic transitions, selection rules, and
photodetectors.

\begin{figure}
\includegraphics[width=0.45\textwidth]{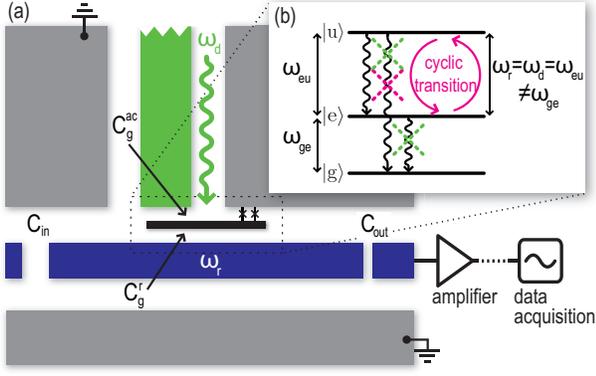}
   \caption{(Color online) Sketch of the mesoscopic shelving qubit readout. 
   (a) A three-level superconducting qubit is capacitively coupled ($C_{\rm g}^{\rm r}$) 
   to a coplanar wave-guide microwave resonator with angular frequency $\omega_{\rm r}$ 
   and input and output capacitors $C_{\rm in}$ and  $C_{\rm out}$, respectively. The 
   qubit is also coupled to an orthogonal transmission line via $C_{\rm g}^{\rm ac}$. (b) 
   The transition $\omega_{\rm eu}$ is resonant to the cavity and is driven 
   with a transversal coherent field (magenta line). Transition rates 
   $|{\rm e}\rangle\rightarrow|{\rm g}\rangle$ and $|{\rm u}\rangle\rightarrow|{\rm g}\rangle$ 
   are reduced by Purcell effect (green dashed lines) or selection rules (magenta dashed lines).}
\label{shelvingsketch}
\end{figure}

In the following, we present a method for implementing a
single-shot QND-like fast high-fidelity readout of superconducting qubits. It
preserves the spirit of electron shelving, but it is suitably
adapted to existent microwave technology in circuit QED where,
for example, single-photon detectors are unavailable.
We assume that the qubit is prepared
in an unknown pure state and that our task is to measure the spin
operator $\sigma^z$.
We consider a three-level superconducting
qubit~\cite{paspalakis04,nori05} inside an on-chip microwave
resonator (acting as a cavity), as shown in Fig.~\ref{shelvingsketch}.
The initial qubit  state is encoded in the two lower
energy levels, $| \psi(0) \rangle = \alpha | {\rm g}
\rangle+ \beta | {\rm e}\rangle$. In
addition, we consider an anharmonic three-level qubit where the
transition frequencies are different, $\omega_{\rm ge} \neq
\omega_{\rm eu}$. Levels $| {\rm e} \rangle$ and $| {\rm u}
\rangle$ are coupled resonantly to a resonator mode,
but there is no dynamics because the resonator is initially empty and
level $| {\rm u}\rangle$ unpopulated. To start with the readout
process, we drive the transition between levels $| {\rm e}\rangle$
and $| {\rm u}\rangle$ with a coherent resonant field with angular
frequency $\omega_ {\rm d}$ and amplitude $\mu$ transversal to the
resonator~axis.

The system Hamiltonian, in the energy eigenbasis and after a
rotating-wave approximation, can be written as
\begin{eqnarray}
H & = & \frac{\hbar\omega_{\rm eu}}{2} \sigma^z_{\rm eu}
       + \hbar\omega_{\rm r} a^{\dagger} a
       + \hbar g_{\rm eu} \left( \sigma_{\rm eu}^+ a
       + \sigma_{\rm eu}^- a^{\dagger}\right)       \nonumber\\
   && + \hbar \Omega_{\rm eu} \left( \sigma_{\rm eu}^+ e^{-i\omega_{\rm d}t}
           + \sigma_{\rm eu}^-  e^{i\omega_{\rm d}t} \right) \nonumber\\
 && + \hbar\lambda\left( a^{\dagger}e^{-i\omega_{\rm d}t}
       +a  e^{i\omega_{\rm d}t}\right).
\label{hamsimple}
\end{eqnarray}
Here, $\sigma^z_{\rm eu} \equiv | {\rm u} \rangle \langle {\rm u}
| - | {\rm e} \rangle \langle {\rm e} |$, $\sigma^+_{\rm eu} \equiv |
{\rm u} \rangle \langle {\rm e} |$, $\sigma^-_{\rm eu} \equiv |
{\rm e} \rangle \langle {\rm u} |$, $a\:(a^{\dagger})$ are the bosonic
annihilation (creation) operators of the resonator field, and $g_{\rm eu}$,
$\Omega_{\rm eu}$, and $\lambda$ are coupling strengths. $\lambda$ describes the crosstalk between the driving field and
the resonator, and its origin typically depends on the specific
setup~\cite{mariantoni08}. The qubit readout happens under the
resonant condition $\omega_{\rm r}=\omega_{\rm d}=\omega_{\rm
eu}$. We assume that the transition $|{\rm u} \rangle \rightarrow
| {\rm e} \rangle$ is sufficiently long-lived such that it does
not decay during the short operation time. Finally, our model
considers enough energy anharmonicity so that the radiative decay
rates associated with the transitions $|{\rm e} \rangle
\rightarrow | {\rm g} \rangle$ and $| {\rm u} \rangle\rightarrow|
{\rm g} \rangle$ are reduced by Purcell effect. Also, these transitions
can be reduced exploiting the characteristic selection rules and symmetry
breaking properties of superconducting qubits~\cite{nori05,deppe08}.

We rewrite the Hamiltonian in a reference frame rotating with the
driving field frequency via the transformation $U^{\rm
rot} = \exp\left[-i\omega_{\rm d}(a^{\dagger}a+\sigma_{\rm
eu}^+\sigma_{\rm eu}^-)t\right]$, obtaining
\begin{eqnarray}
   H^{\rm rot} \! = \hbar  \Omega_{\rm eu} \sigma_{\rm eu}^x
       \! + \! \hbar g_{\rm eu} \left( \sigma_{\rm eu}^+ a + \sigma_{\rm eu}^-  a^{\dagger} \right) \! + \! \hbar \lambda \left( a^{\dagger}+  a \right) ,
\end{eqnarray}
with $\sigma_{\rm eu}^x =\sigma_{\rm eu}^+ +\sigma_{\rm eu}^- $.
We now apply the transformation $U^{I}=\exp\left[- i
\Omega_{\rm eu} \sigma_{\rm eu}^x t \right]$ under the strong-driving condition
   $ \Omega_{\rm eu} \gg g_{\rm eu}$~\cite{solano03}, and derive the effective Hamiltonian
\begin{eqnarray}
\label{eq:heff}
   H_{\rm eff}=\frac{\hbar g_{\rm eu}}{2} ( \sigma_{\rm eu}^+ + \sigma_{\rm eu}^- )
       (a+a^{\dagger}) + \hbar \lambda \left( a^{\dagger}+  a
       \right) \, .
\end{eqnarray}
The first part of the Hamiltonian simultaneously realizes
Jaynes-Cummings and anti-Jaynes-Cummings resonant interactions. It
does not generate Rabi oscillations, but conditional field
displacements~\cite{solano03}, while the second term implements
a resonant displacement. The initial qubit--field state is
$| \psi(0) \rangle = \alpha|\,{\rm g} \,\rangle | 0 \rangle +
\beta \left( |+\rangle+|-\rangle\right) |0\rangle / \sqrt{2}$,
with $\sigma_{\rm eu}^x | \pm \rangle =  \pm | \pm \rangle$. After
an interaction time $t$, the state is
\begin{eqnarray}
\label{eq:evolved}
   | \psi(t) \rangle & = & \alpha | {\rm g} \rangle | \bar{\nu}(t)\rangle
       + \frac{\beta}{\sqrt{2}} \big\lbrack |+\rangle|\bar{\eta} (t)
           + \bar{\nu}(t)\rangle \nonumber \\
       && \qquad \qquad \qquad + | -\rangle | - \bar{\eta} (t)
           +\bar{\nu} (t) \rangle \big\rbrack .
\end{eqnarray}
Here, the coherent states
$|\,\pm\bar{\eta}(t)+\bar{\nu}(t)\rangle$, with $\bar{\eta}(t) = -i g_{\rm eu}t/2$
and $\bar{\nu}(t) = -i \lambda t$, are generated by the
displacement operators
$\mathcal{D}[\pm\bar{\eta}(t)+\bar{\nu}(t)]=\exp
\left([\pm\bar{\eta}(t)+\bar{\nu}(t)] a^{\dagger}-[\pm {\bar
\eta}^{\ast}(t)+{\bar \nu}^{\ast}(t)] a\right)$. In general, we
expect the crosstalk to be small, so that $\lambda
\ll g_{\rm eu} / 2$ and $\bar{\nu}(t) \ll \bar{\eta}(t)$. When the
measurement starts, the applied driving field yields many
intracavity photons ${\bar N}^{\rm e}_{\rm in} (t) \approx |
\bar{\eta}(t) |^2$ with probability $| \alpha |^2$ if the state $|
{\rm e} \rangle$ is projected. If the state $| {\rm g} \rangle$ is
selected, it yields a few photons ${\bar N}^{\rm g}_{\rm in} (t) =
| \bar{\nu}(t) |^2 \ll | \bar{\eta}(t) |^2$ with probability $|
\beta |^2$. 

We now add to our model a zero-temperature dissipative
reservoir for the cavity field, characterized by a decay rate $\kappa$. The
corresponding master equation reads
\begin{eqnarray}
\label{masterequation}
   \dot{\rho}_{\rm q-f} =-\frac{i}{\hbar}[H_{\rm eff},\rho_{\rm q-f}]
       +\mathcal{L}_{\rm f} \, \rho_{\rm q-f}\ \ ,
\end{eqnarray}
with $\mathcal{L}_{\rm f} \rho_{\rm q-f} \equiv \kappa L [ a ] \rho_{\rm q-f}$ such that
\begin{eqnarray}
\mathcal{L}_{\rm f} \rho_{\rm q-f} = \kappa \left( 2 a \rho_{\rm q-f} a^{\dagger} - a^{\dagger} a\rho_{\rm q-f} - \rho_{\rm q-f} a^{\dagger}a \right)/2
\end{eqnarray}
and expansion $\rho_{\rm q-f} (t) = \sum_{ j , k  = {\rm g} , - , + }
| j \rangle \langle k | \otimes \rho^{j  k}_{\rm f}(t)$. Here, it is possible to find analytical solutions for $\rho^{j
k}_{\rm f}(t) = \langle j | \rho_{\rm q-f} (t) | k \rangle$ using
standard phase-space tools~\cite{lougovski} and the method of
characteristics to solve the partial differential
equations~\cite{barnett}. The solutions read
\begin{eqnarray}
\label{eq:rhoend}
   \rho^{+ +}_{\rm f}(t) & = & \frac{|\beta|^2}{2}|\eta(t)+\nu(t) \rangle
       \langle\eta(t)+\nu(t) |,                                    \nonumber \\
   \rho^{- -}_{\rm f}(t) & = & \frac{|\beta|^2}{2}|-\eta(t)+\nu(t) \rangle
       \langle -\eta(t)+\nu(t) |,                          \nonumber \\
   \rho^{{\rm g g}}_{\rm f}(t) & = & |\alpha|^2 |\nu(t)\rangle\langle \nu(t)| ,
   \nonumber \\
   \rho^{+ -}_{\rm f}(t) & = & \frac{|\beta|^2}{2}\frac{f_1(t)}{e^{-2|\eta(t)|^2}}
       |\eta(t)+\nu(t)\rangle\langle -\eta(t)+\nu(t) |,    \nonumber \\
   \rho^{{\rm g} +}_{\rm f}(t) & = & \frac{\beta^{\ast}\alpha}{\sqrt{2}}
       \frac{f_2(t)}{e^{-|\eta(t)|^2/2}}
       |\nu(t) \rangle\langle \eta(t)+\nu(t)|,                 \nonumber \\
   \rho^{{\rm g} -}_{\rm f}(t) & = & \frac{\beta^{\ast}\alpha}{\sqrt{2}}
       \frac{f_2(t)}{e^{-|\eta(t)|^2/2}}
       |\nu(t) \rangle\langle -\eta(t)+\nu(t)| ,
\end{eqnarray}
where
\begin{eqnarray}
 f_1(t) & = & \exp\left( -2\frac{g_{\rm eu}^2}{\kappa}t
   + \frac{4g_{\rm eu}^2}{\kappa^2}\left[1-e^{-\kappa t/2}\right] \right), \nonumber\\
 f_2(t) & = & \exp\left( - \frac{g_{\rm eu}^2}{2\kappa}t
   + \frac{g_{\rm eu}^2}{\kappa^2}\left[1-e^{-\kappa t/2}\right] \right) ,
\end{eqnarray}
with $\eta(t) = -i g_{\rm eu}/\kappa \left[1-e^{-\kappa t/2}\right]$,
$\nu(t) = -2i\lambda/\kappa  \left[1-e^{-\kappa t/2}\right]$.
For a small crosstalk $\lambda$, the leakage rate of outgoing 
photons $N^{\rm e}_{\rm out} (t) $ when state $| {\rm e} \rangle$ is measured can be estimated as
\begin{eqnarray}
\label{analyticalphotons}
   N^{\rm e}_{\rm out} (t) = \kappa \, N^{\rm e}_{\rm in} (t)
       = \kappa \, | \eta (t) |^2
       = \frac{g_{\rm eu}^2}{\kappa} \left(1-e^{-\kappa t/2}\right)^2  ,
\end{eqnarray}
where $N^{\rm e}_{\rm in}(t) =  | \eta (t) |^2$ is the intracavity mean photon number.  $N^{\rm e}_{\rm out} (t)$ grows very fast well below decoherence times. It can be measured,
e.g., by means of a data acquisition card, which follows a
phase-preserving or even a more quiet
phase-sensitive~\cite{phase:sensitive:amps} linear amplifier. We
also notice that one can profit from the generated large intracavity field to adapt to other readout techniques~\cite{hofheinz08}.

\begin{figure}
\includegraphics[width=0.45\textwidth]{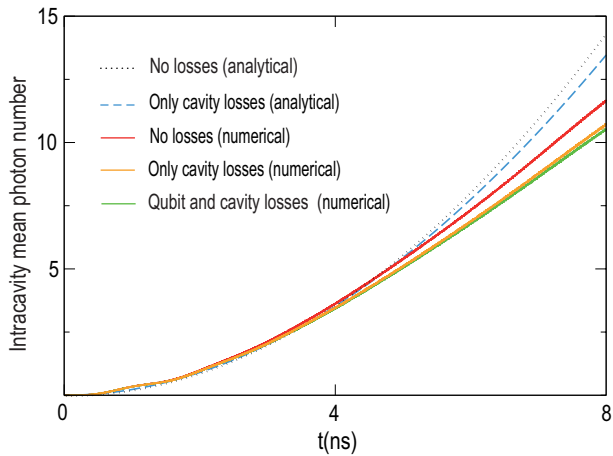}
\caption{(Color online) Intracavity mean photon number for the mesoscopic shelving readout of a CPB in the charge-phase regime with conservative parameter set: $E_{\rm C}/\hbar=E_{\rm J}/\hbar=2\pi\times 10\,\rm{GHz}$, $g_{\rm eu} =\Omega_{\rm eu}/5 = 2 \pi \times 150\,\rm{MHz}$, $\kappa = 2 \pi \times 1.6 \, \rm{MHz}$,
$\gamma_{\rm eu}=10\gamma_{\rm ge}=2\,\rm{MHz}$,
$\gamma_{\rm gu}=0$, and $\gamma_{\varphi}= 2\,\rm{MHz}$. The dotted and dashed curves correspond to the analytical results and the solid lines to the numerical results. Note that, in the absence of losses, there is still a difference between the analytical and the numerical results. This is due to the off-resonant couplings and multilevel character of the realistic model.}
\label{shelvingsimulation}
\end{figure}

The physical concepts behind the mesoscopic shelving are general and can be adapted to different qubits and setups. We exemplify here with a possible adaptation to a Cooper-pair box (CPB) coupled to a microwave resonator of angular frequency $\omega_{\rm
r}$. Here, the CPB has a Josephson energy $E_{\rm J}$
and charging energy $E_{\rm C}=(2e)^2/2C_{\rm tot}$, where $C_{\rm tot}$ is the total island capacitance. We refer to a system that is essentially the one in
Ref.~\cite{blais04}, with the addition of a transmission line, orthogonal to the resonator, for driving the qubit (see
Fig.~\ref{shelvingsketch}). Using $n$, the number operator for
excess Cooper pairs on the CPB island, and $\varphi$, the phase difference across the Josephson junction, the Hamiltonian can be written as
\begin{eqnarray}
\label{hamilton-charge}
H &=& E_{\rm C}(n-n_x)^2-E_{\rm J}\cos{\varphi}
          + \hbar\omega_{\rm r}a^{\dagger}a \ ,
\end{eqnarray}
where
\begin{eqnarray}
(2e) n_x & \!\! = \!\! & C^{\rm dc}_{\rm g} V^{\rm dc}_{\rm g}+C^{\rm ac}_{\rm g} V^{\rm ac}_{\rm g}(t)
       + C^{\rm r}_{\rm g} V_0(a^{\dagger}+a)  \, .
\end{eqnarray}
Here, $C^{l}_{\rm g}$ ($l=\{{\rm dc,ac,r}\}$) are effective
gate capacitances, $V_{\rm g}^{\rm dc}$ is the gate voltage that
defines the working point (we choose the so-called ``sweet spot''
$C^{\rm dc}_{\rm g}V_{\rm g}^{\rm dc}/2e=1/2$), $V_{\rm g}^{\rm
ac}$ is the voltage of the orthogonal driving field, $a$
($a^{\dagger}$) refers to the cavity field, and $V_0$ is
the resonator zero-point voltage. Note that the CPB is coupled to
the resonator and the ac drive via the charge number operator. The
classical gate charge $C_{\rm g}^{\rm ac}V_{\rm g}^{\rm ac}$ and
the quantum gate charge $C_{\rm g}^{\rm r}V_0$ represent small
deviations from the sweet spot.

We can rewrite $H$ in a basis of CPB
eigenstates restricted to the first three energy levels, the ground state $| {\rm g} \rangle$
and the first and second excited states, $|{\rm e}\rangle$ and $|{\rm u}\rangle$, respectively.
This leads to an effective Hamiltonian for the driven qubit--resonator
system,
\begin{eqnarray}
\label{hamilton-eigen}
 {\cal H} & = & {\cal H}_0 + {\cal H}_{\rm int} + {\cal H}_{\rm d} \ ,  \\
   {\cal H}_0 &=& \sum_{j = {\rm g,e,u}} E_j |j\rangle \langle j|
       + \hbar\omega_{\rm r} a^{\dagger}a          \ ,         \nonumber\\
   {\cal H}_{\rm int} &=& \hbar(g_{\rm gu}|{\rm u}\rangle\langle {\rm g}|
           + g_{\rm eu}|{\rm u}\rangle \langle {\rm e}|)(a^{\dagger}+a)
                           \, + \, {\rm H.c.} \ ,          \nonumber\\
   {\cal H}_{\rm d} &=&  (\cos{\omega_{\rm d}t})
     \sum_{j<k= {\rm g,e,u}}(\hbar\Omega_{jk}| j \rangle \!\langle k|
       \, + \, {\rm H.c.})     \ .     \nonumber
\end{eqnarray}
The coupling strengths
$g_{jk}\equiv(E_{\rm C}/\hbar e) n_{jk} C_{\rm g}^{\rm r}V_0$ and
$\Omega_{jk}\equiv(E_{\rm C}/\hbar e) n_{jk} C_{\rm g}^{\rm ac}V^{\rm ac}_{\rm g}(t)$
are proportional to the matrix elements
$n_{jk}=\langle k | n | j \rangle$. In order to obtain the time evolution of the complete system, including the relaxation and dephasing of qubit transitions, we
numerically solve the master equation for the qubit-resonator density matrix $\rho(t)$
\begin{eqnarray}
\dot{\rho} & = & - \frac{i}{\hbar} [ {\cal H} , \rho ] +
{\mathcal{L}}_{\rm f} \rho + {\mathcal{L}}_{\rm q} \rho  \ ,
\label{fulldyn}
\end{eqnarray}
where, using the functional $L$ defined in Eq.~(\ref{masterequation}), we have
\begin{eqnarray}
{\mathcal{L}}_{\rm q} \rho
 & \!\! = \!\! & \sum_{j<k} \gamma_{jk} L \big[ \, | k \rangle\langle j| \, \big] \rho
+\sum_{j} \frac{\gamma_{\varphi}}{2} L \big[ \, |j \rangle\langle j| \, \big] \rho \, .
\end{eqnarray}
Here, $\kappa$ is the decay rate of the resonator, $\gamma_{jk}$ $(j,k=\{{\rm g,e,u}\})$
are the relaxation rates for the transitions $| {\rm e} \rangle \rightarrow | {\rm u} \rangle$ and
$| {\rm u} \rangle \rightarrow | {\rm g} \rangle$, and $\gamma_{\varphi}$ is the dephasing rate, which
we take to be equal for all coherences.
In the numerical solution of Eq.~(\ref{fulldyn}), we truncate the resonator Hilbert space
to $25$ photon number states  due to technical limitations. In addition,
we make sure that the population of the fourth
qubit eigenstate is negligible.
Clearly, the condition
$\gamma_{\rm eu}+\gamma_{\rm gu}\ll g_{\rm eu}$
is crucial for our method to be efficient. Therefore, a qubit layout with suppressed $\gamma_{\rm gu}$
is preferred for the shelving readout.

The results for the intracavity mean photon number with a conservative set of parameters, in the analytical and numerical cases, are shown in Fig.~\ref{shelvingsimulation}. We see that, although the full dynamics in Eq.~\eqref{fulldyn} is considerably more complex than the one in Eq.~\eqref{masterequation}, the simple analytical model captures the
essence of the system dynamics. The main influence of a realistic description is a small reduction in the intracavity mean photon number. We observe that, given the short interaction times displayed in Fig.~\ref{shelvingsimulation}, the resonator decay rate $\kappa$ alone has a small effect in the cavity population, while the finite lifetime of states $|{\rm e}\rangle$ and $|{\rm
u}\rangle$ is slightly more important. In this manner, we feel comfortable to extrapolate the analytical results for the cavity population including cavity losses for short measurement times to make further estimations.

The signal-to-noise ratio (SNR) after a measurement time
$\tau_{\rm m}$ is the ratio between the accumulated number of outgoing photons and the accumulated noise~\cite{blais04}. The latter is dominated by the amplifier noise, $n_{\rm
amp}=k_{\rm B}T_{\rm n}/\hbar\omega_{\rm r}\simeq 25$, where
$T_{\rm n}$ is its associated noise temperature. In this manner,
\begin{equation}
       \mathrm{SNR}(\tau_{\rm m})=\frac{\int_0^{\tau_{\rm m}}\kappa N_{\rm in}^{\rm e}(t) dt}
                              {n_{\rm amp} B \tau_{\rm m}}
\end{equation}
where $B \equiv \max{\{\kappa,\gamma_{jk}\}}$ is the measurement
bandwidth. We now estimate the SNR for three relevant consecutive
times.
First, we use the maximum simulated time $\tau^{\rm sim}_{\rm m}\simeq 8 \,\rm{ns}$, 
corresponding to $10$ intracavity photons (cf. 
Fig.~\ref{shelvingsimulation}). We obtain a ${\rm SNR} \simeq 0.2$. 
Considering that our simulations include all relevant system
details without any approximation~\cite{boissonneault}, this is a
remarkable result for such an extremely short measurement
time. Using our analytical results including resonator
dissipation, see Eq.~(\ref{analyticalphotons}), we estimate that a critical measurement time
$\tau^{\rm crit}_{\rm m} \simeq 19$\,ns is necessary to reach the
condition ${\rm SNR} \sim 1$. This is the minimum time required for a
single-shot measurement of the qubit state $|{\rm e}\rangle$.
Finally, to achieve high-fidelity qubit readout, we choose the measurement time $\tau^{\rm hf}_{\rm m} \simeq
50$\,ns which
corresponds to ${\rm SNR} \simeq 6.2$ and fidelity $F = 1 - e^{\rm -SNR} = 0.998$. Notably, $\tau^{\rm hf}_{\rm
m} \ll 1/\gamma_{\rm ge}$. Consequently, we expect a single-shot measurement of the qubit state $|{\rm e}\rangle$ with fidelities close to $1$. The proposed mesoscopic shelving qubit readout is of a QND-like character, due to the continuous cavity field amplification in each measurement event. In addition, $\tau^{\rm hf}_{\rm m}$ is at least one order of magnitude shorter than typical measurement times employed in the state-of-the-art experiments based on dispersive readouts. Note that, even for a driving-resonator crosstalk of $\lambda = 2 \pi \times 10\,\rm{MHz}$~\cite{mariantoni08}, the cavity population associated with the measurement of state $|{\rm g}\rangle$ is well below the amplifier background noise level.

In summary, we have presented a novel qubit 
readout scheme based on a mesoscopic shelving technique, allowing a fast high-fidelity single-shot QND-like measurement of superconducting qubits in circuit QED. 

We acknowledge stimulating discussions with
P.~Bertet, M.~Hofheinz, A. Wallraff, J.~M.~Martinis, R.~Schoelkopf, R.~Bianchetti, and F.~Deppe. This work is
funded by Deutsche Forschungsgemeinschaft through SFB 631, Heisenberg Programme, German Academic Exchange Service,
and German Excellence Initiative via the Nanosystems
Initiative Munich (NIM). E.S. thanks Ikerbasque Foundation, UPV-EHU Grant GIU07/40, and EuroSQIP European project.

\end{document}